\documentclass[twocolumn,showpacs,prl,aps, longbibliography]{revtex4-2}
\usepackage{amsfonts}
\usepackage{amsmath}
\usepackage{physics}
\usepackage{graphicx}
\usepackage[unicode]{hyperref}
\hypersetup{
   unicode=true,          %
   plainpages=false,
   colorlinks=true,       
   citecolor=blue,        
   urlcolor=blue
}
\usepackage[caption=false,position=top,singlelinecheck=off,justification=raggedright]{subfig}
\usepackage{color}

\begin{document}

\title{Universal logic gates for coupled period-doubling systems}

\author{Emmanuel D.G. U}
\email[]{edu@up.edu.ph}
\affiliation{National Institute of Physics, University of the Philippines, Diliman, Quezon City 1101, Philippines}

\author{Roy D. Jara Jr.}
\affiliation{National Institute of Physics, University of the Philippines, Diliman, Quezon City 1101, Philippines}

\author{Jayson G. Cosme}
\email[]{jcosme@nip.upd.edu.ph}
\affiliation{National Institute of Physics, University of the Philippines, Diliman, Quezon City 1101, Philippines}

\date{\today}

\begin{abstract}

We propose a general architecture for universal logic operations using NAND and NOR gates on classical information encoded in period-doubled states of periodically-driven systems. The protocol involves applying a single pulse that simultaneously couple two input nodes with an output node. We show that the states of the nodes can be precisely controlled by tuning the coupling strength and pulse duration, allowing for robust logic gate operation. To highlight the universality of the protocol, we demonstrate its applicability on different systems, such as classical networks of dissipative parametric oscillators (DPO), quantum networks of Kerr parametric oscillators (KPO), and the periodically-driven open Dicke lattice model (DLM) emulating discrete time crystals (DTCs). We identify the parameter regimes in which the logic gate architecture is valid, and we showcase its robustness in the presence of fluctuations.

\end{abstract}

\maketitle

\begin{figure}
    \centering
    \includegraphics[scale = 0.332]{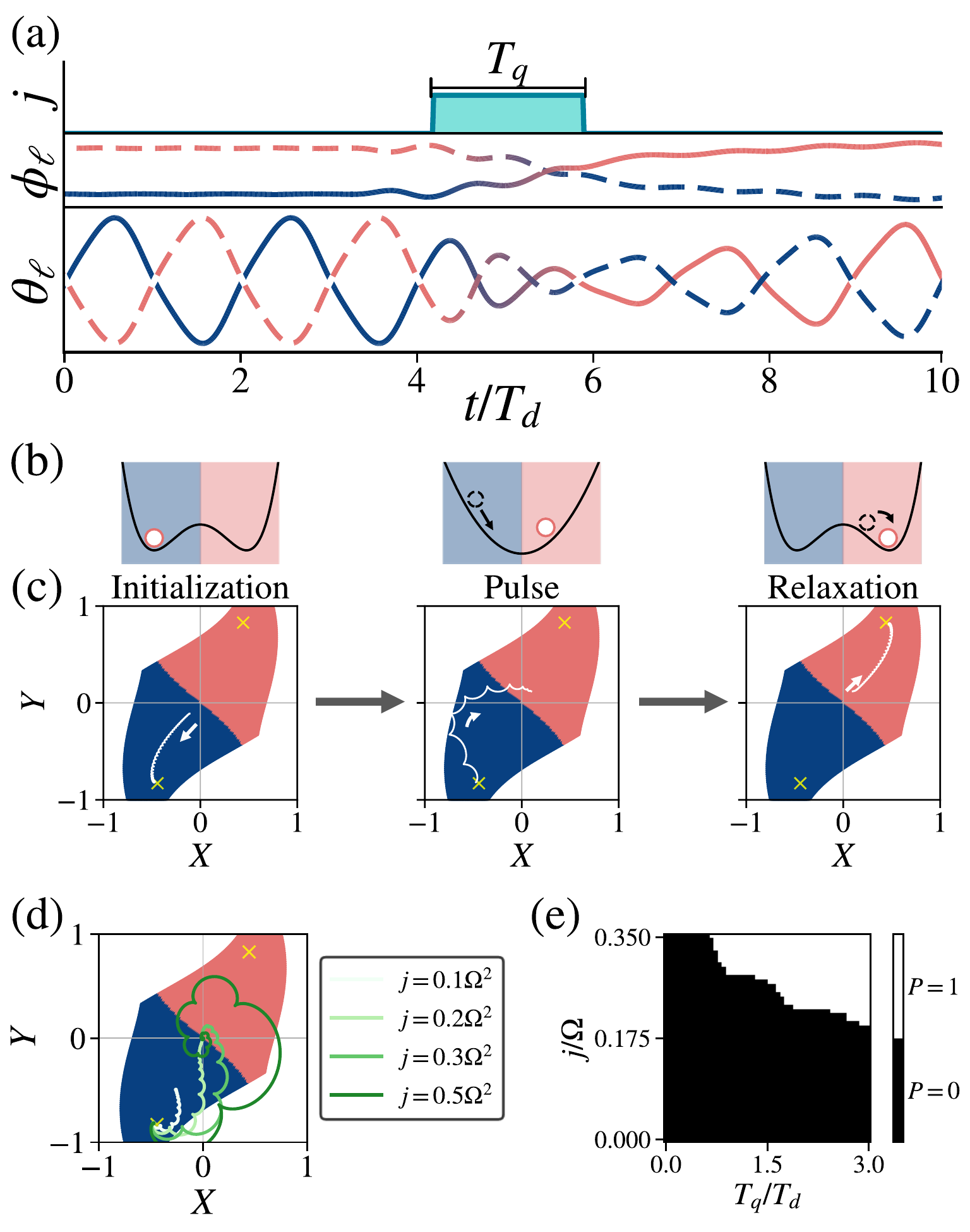}
    \caption{(a) (Top panel) The coupling strength $j$ between the two oscillators. (Middle panel) The absolute time phase $\phi_\ell$ and (Bottom panel) Exemplary dynamics of $\theta_\ell$ for the two oscillators, demonstrating how the rapid coupling leads to a bit flip operation. The dashed (solid) lines represents the first (second) site. (b) Sketch of the potential landscape of the system and (c) dynamics of the system in the rotating-frame phase space during the (Left) initialization, (Middle) pulse, and (Right) relaxation periods. The dark-shaded (light-shaded) region corresponds to the basin of attraction for the 1-bit (0-bit), while the cross marker indicates the location of their respective period-doubled state. (d) The trajectory of a DPO in phase space after being coupled indefinitely to another in-phase DPO for various $j$. (e) Phase diagram showing when the bit flip operation succeeds as a function of $j$ and $T_q$. The parameters used are $\{ \Omega_d, A, \gamma\} = \{ 2.0\Omega, 0.5, 0.2\Omega \}$.}
    \label{fig:fig1}
\end{figure}

When driven near resonance, a DPO can break the discrete time translation symmetry associated with the external drive, spontaneously selecting from the two degenerate period-doubled states in Fig.~\ref{fig:fig1}(a) \cite{kovacic_2018_poperioddouble}. This event is signaled by persistent oscillations at a subharmonic frequency of the drive and the exponential growth of the oscillator's response amplitude, saturating at a constant value in nonlinear systems. The robustness of these period-doubled states against thermal fluctuations along with their ubiquity across a wide range of physical systems \cite{leuch_2016_posymmbreaking, apffelfleury_2024_topologicalpo, bello_persistent_2019, fabiani_2022_pobits, mahboob_2008_pobits, nosaneichler_2019_pobits, elyasi_2022_pobits, frimmer_2019_rapidflippingpo} makes them strong candidates for storing information as bits, motivating various proposals for performing logic operations on parametric oscillator networks \cite{thegoto_1959_poqubits, hiroya_2021_pologic, masuda_2021_pologic, hatanaka_2017_pologic, linyamamoto_2014_pologic}.

The breaking of discrete time translation symmetry has also been observed in many-body systems, leading to the formation of a DTC \cite{yao_2017_dtcs, else_2020_dtcreview1, zaletel_2023_dtcreview2, else_2016_dtcs, lazarides_2020_theodtc}. Thus far, DTCs have been theoretically and experimentally investigated in many different platforms, such as superconducting qubits \cite{frey_2022_superconductingdtcs}, spin systems \cite{sullivan_2020_expdtc, zhang_2017_experimentaldtc1, gambetta_2019_theodtc, khemani_2016_dtcs, chan_2015_spindtc, frey_2022_spindtc, owen_2018_spindtc, krishna_2023_spindtc, munozarias_2022_spindtc, russomanno_2017_spindtc, keyserlingk_2016_spindtc}, atom-cavity systems, and other platforms \cite{liu_2024_rydbergdtcs, sacha_2015_dtcs, heugel_2019_classicaldtcs, yao_2020_classicaldtcs}. Here, we focus specifically with period-doubled DTCs, which manifest as a period-doubled response, producing two degenerate, symmetry-broken states akin to those shown in Fig.~\ref{fig:fig1}(a). As demonstrated in Refs.~\cite{roy_2024_parametricmapping, poletti_2018_perioddoublingdtcs} certain class of systems with period-doubled DTCs can be effectively treated as parametric oscillators. Their DTC states can then be used as classical bits and perform simple controlled operation like bit-flips \cite{roy_2025_atp}. This correspondence motivates the question of whether there exists a set of driving protocols for performing universal logical gate operations on DTCs.

In this work, we propose a general protocol for realizing NAND and NOR gates \cite{kotb_2024_nanduniversal} for classical bits encoded in the period-doubled states of DPOs. To this end, we first consider the dynamics of two coupled DPOs as a minimal model. In the representative dynamics shown in Fig.~\ref{fig:fig1}(a), we demonstrate that applying a short pulse between two coupled DPOs can flip the oscillation phase of each oscillator, owing to the system's nonlinear response. This coupling modifies the phase space landscape of the combined system, as sketched in Fig.~\ref{fig:fig1}(b), which then leads to a nonlinear trajectory away from the original symmetry-broken state. As shown in Fig.~\ref{fig:fig1}(c), a bit-flip operation or a NOT gate can be implemented by switching the coupling off once the system is pushed into the basin of attraction of the opposite symmetry-broken state. We use this principle to propose a scheme for performing NAND and NOR operations by employing a single pulse protocol that simultaneously couples three DPOs, which is the minimal number of nodes for a universal logic. We demonstrate the validity of our proposed architecture on three periodically-driven systems: a classical network of DPOs, three-site Kerr parametric oscillators (KPOs), and the Dicke lattice model (DLM). With these three systems, we verify high fidelity on the performance of the logic operation across a wide range of parameter regimes, even in the presence of fluctuations. Our proposed scheme leverages the absolute time phase of the oscillations as a measurable quantity for tracking trajectories in phase space. Thus, we provide a simple yet effective method for performing universal logic operations in any generic periodically-driven system exhibiting parametric resonance.

\textit{Dissipative parametric oscillators --} To describe a network of coupled DPOs, we consider a system of linearly coupled pendulums connected to a thermal bath. This system is described by the equation
\begin{equation}\label{eq:sg_eom}
    \ddot{\theta}_\ell + \Omega^2[1-A\cos (\Omega_d t)]\sin \theta_\ell + \gamma \dot{\theta}_\ell - j(\theta_{\ell-1} + \theta_{\ell + 1}) = \eta_\ell
\end{equation}
where $\theta_\ell(t)$ is the angle the $\ell$-th oscillator relative to the vertical axis, $\Omega$ is the oscillator's natural frequency, while $A$ and $\omega_{d}$ are the driving strength and frequency respectively. The dissipation rate is set by $\gamma$, while $j$ sets the coupling strength between nearest-neighbor oscillators. The Gaussian white noise $\eta_\ell$ describes the effect of thermal noise on the system, and it satisfies $\langle \eta_\ell (t)\rangle = 0$ and $\langle \eta_\ell (t) \eta_{\ell'} (t')\rangle = 2 \widetilde{T} \Omega^2 \gamma \delta{(t - t')} \delta_{\ell\ell'}$, where $\widetilde{T} = k_B T / (mL^2\Omega^2)$ \cite{yao_2020_classicaldtcs}.

When driven near-resonance, $\omega_{d} \approx 2 \Omega$, the DPO exhibits a period-doubling instability \cite{kovacic_2018_poperioddouble, leuch_2016_posymmbreaking}, spontaneously choosing between the two symmetry-broken states in Fig.~\ref{fig:fig1}(a), which we use to represent a classical bit. To distinguish between the two period-doubled states, we calculate the absolute time phase $\phi(t)$, which can be extracted from $\theta_\ell$ according to 
\cite{simula_2024_atp, roy_2025_atp}

\begin{align}\label{eq:atp}
    \theta_{\ell, R} (t) = r_\ell(t) e^{i \phi_\ell(t)} = \frac{\omega_R}{\pi} \int_t^{t+\frac{2\pi}{\omega_R}} e^{i\omega_R \tau} \theta_\ell(\tau) d\tau,
\end{align}
where $\theta_{\ell, R}$ is the complex amplitude associated with the period-doubled response with of the $\ell$-th site with frequency $\omega_R$. Here, we denote the state with $\phi_\ell < 0$ ($\phi_\ell > 0$) as the 1-bit (0-bit). Note that due to the degeneracy of the period-doubled states, this choice of label is completely arbitrary. We also extract the state of the $\ell$-th site in the rotating-frame phase space spanned by $X = \mathrm{Re}[\phi_\ell(t)]$ and $Y = \mathrm{Im}[\phi_\ell (t)]$.

In Fig.~\ref{fig:fig1}(a), we present a simple principle for switching the state of the two DPOs by tracking their phase space trajectory using the absolute time phase $\phi_\ell$, as depicted in Fig.~\ref{fig:fig1}(b) and \ref{fig:fig1}(c). First, the two sites are connected for a finite duration $T_q$, chosen such that the phase space trajectory is pushed into the region of the opposite basin of attraction. These basins, marked by the shaded regions, are associated with the two period-doubled states, represented by the cross markers. Details for the construction of the basins are included in~\cite{suppmat}. Due to the coupling, the system's effective potential becomes modified. This shifts the location of the two symmetry-broken states, thus pushing the system onto a nonlinear trajectory across the phase space. Once the coupling is switched off, if the system is in the other basin right after the coupling, it eventually relaxes towards the symmetry broken partner of the initial state. This sequence effectively implements a bit-flip operation on both DPOs. 

To determine the optimal $T_{q}$ and $j$ that induces a bit-flip, we consider a scenario in which the two DPOs are initialized in the same bit state and are subsequently coupled for the remainder of the simulation. We display in Fig.~\ref{fig:fig1}(d) the trajectory of the first site in the rotating-frame phase space after the coupling is activated for various $j$. For $j = 0.1\Omega^2$, the coupling simply pushes the system into a period-doubling state with a smaller amplitude. Increasing the coupling strength further decreases the oscillation amplitude to zero, signifying the decay of period-doubling in coupled DPOs for strong couplings. We can exploit this nonlinear decay to perform a bit-flip, since as the trajectory approaches the new steady-state, it winds around the origin and visits the opposite basin of attraction. This behavior can be seen in the trajectories for $j > 0.2\Omega^{2}$, as shown in Fig.~\ref{fig:fig1}(d). The system's state can then be flipped by turning off the coupling at an appropriate time $T_q$, specifically when the system is located within the opposite basin of attraction, as depicted in the last panel of Fig.~\ref{fig:fig1}(c). We refer to the finite coupling duration as the pulse stage. For the parameters in Fig.~\ref{fig:fig1}(d), we find that the phase space trajectory explores the opposite basin only when $j$ is beyond its critical value. This is consistent with the results displayed in Fig.~\ref{fig:fig1}(e), which shows the values of $j$ and $T_{q}$ for which the bit-flip succeeds ($P=1$) or fails ($P=0$) when the DPOs are initialized in-phase.

\begin{figure}
    \centering
    \includegraphics[scale = 0.245]{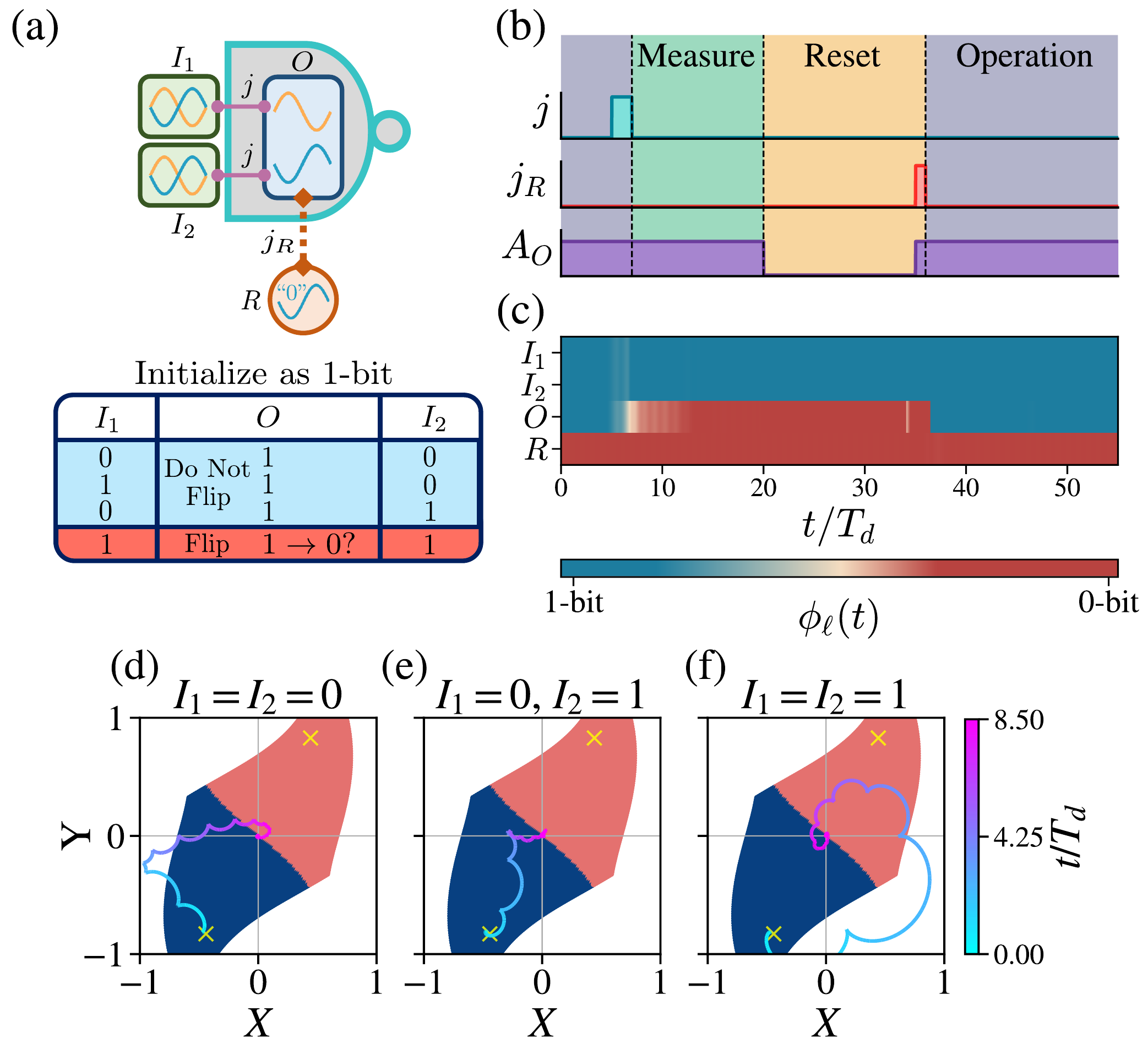}
    \caption{(a) Sketch of the proposed logic architecture. (b) Protocol for the logic operation. (Top panel) The coupling strength $j$ between the output and input sites. (Middle panel) The coupling strength $j_R$ between the output and reference site. (Bottom panel) The driving amplitude $A_\mathrm{O}$ for the output site. (c) Exemplary dynamics showing the absolute time phase $\phi_\ell(t)$ of each site during the logic operation and the reset protocol. (d) - (f) Dynamics of $O$ in the rotating-frame phase space while connected to $I_1$ and $I_2$ for all possible input configurations. The parameters used are $\{ \Omega_d, A, \gamma, j\} = \{ 2.0\Omega, 0.5, 0.2\Omega, 0.3\Omega^2 \}$.}
    \label{fig:fig2}
\end{figure}

\textit{Universal logic operations --} Similar to the case of two coupled DPOs, coupling together three DPOs pushes each site onto a transient, nonlinear path in phase space. This allows us to dynamically control the sites' final state by switching off the coupling after a time $T_q$ when the system falls within the desired basin of attraction. Using our simple framework of timing-based pulse protocols, we now introduce how to perform NAND and NOR operations in a network of four DPOs shown in Fig.~\ref{fig:fig2}(a). In this architecture, two sites $I_1$ and $I_2$ play the role of the input nodes, one site $O$ plays the role of the output node, and one site $R$ acts as a reference node. For a NAND (NOR) gate, the output site must be initialized in the 1-bit (0-bit) state before the operation. To perform the logic operation, a single-pulse is applied to the coupling between the input and output nodes, after which the result is obtained by reading the state of $O$ after it has relaxed to a steady state. The control sequence for the protocol is shown in Fig.~\ref{fig:fig2}(b), while the corresponding exemplary dynamics in Fig.~\ref{fig:fig2}(c) demonstrate a NAND gate operation for the input $I_{1} = I_{2} = 1$. Although only three sites are required to perform a single operation, a fourth site is required to reset the output site to its initial state (1-bit for a NAND gate) in between subsequent operations. As shown in Ref.~\cite{emman_2026_prbpaper}, this can be accomplished by setting $A_\mathrm{O} = 0$, which turns off the parametric driving in the output site, destroying the period-doubled state. A pulse is then applied to the coupling between the output and reference nodes while simultaneously reinstating the driving in the output site. This causes it to anti-synchronize with the reference node. More information regarding the reset protocol can be found in \cite{suppmat}. Note that the NAND logic is equivalent to that of NOR under the transformation $0 \leftrightarrow 1$, which is always valid due to the degeneracy of the period-doubled states representing the bits. As such, we will focus on the NAND gate as the results also apply to the NOR gate.

To choose a suitable pulse duration $T_q$ for a given $j$, we again consider the scenario in which the three DPOs are initialised in a stable period-doubled state before being coupled for the remainder of the simulation. In Figs.~\ref{fig:fig2}(d)-(f), we track the trajectory of the output site in the rotating-frame phase space for all four possible input configurations, given the same driving parameters. The $T_{q}$ is chosen such that the output site enters the opposite basin of attraction of the initial state only for $I_{1} = I_{2} = 1$, in agreement with the NAND truth table shown in Fig.~\ref{fig:fig2}(a). Note that since $I_1$ and $I_2$ are both connected only to $O$, the choice of site label is arbitrary and we can always redefine $I_1 \rightarrow I_2$. Thus, the situation in Fig.~\ref{fig:fig2}(e) for $I_1 = 0, I_2 = 1$ is equivalent to $I_1 = 1, I_2 = 0$.

\begin{figure}\label{fig:Dicke_Validity}
    \centering
    \includegraphics[scale = 0.234]{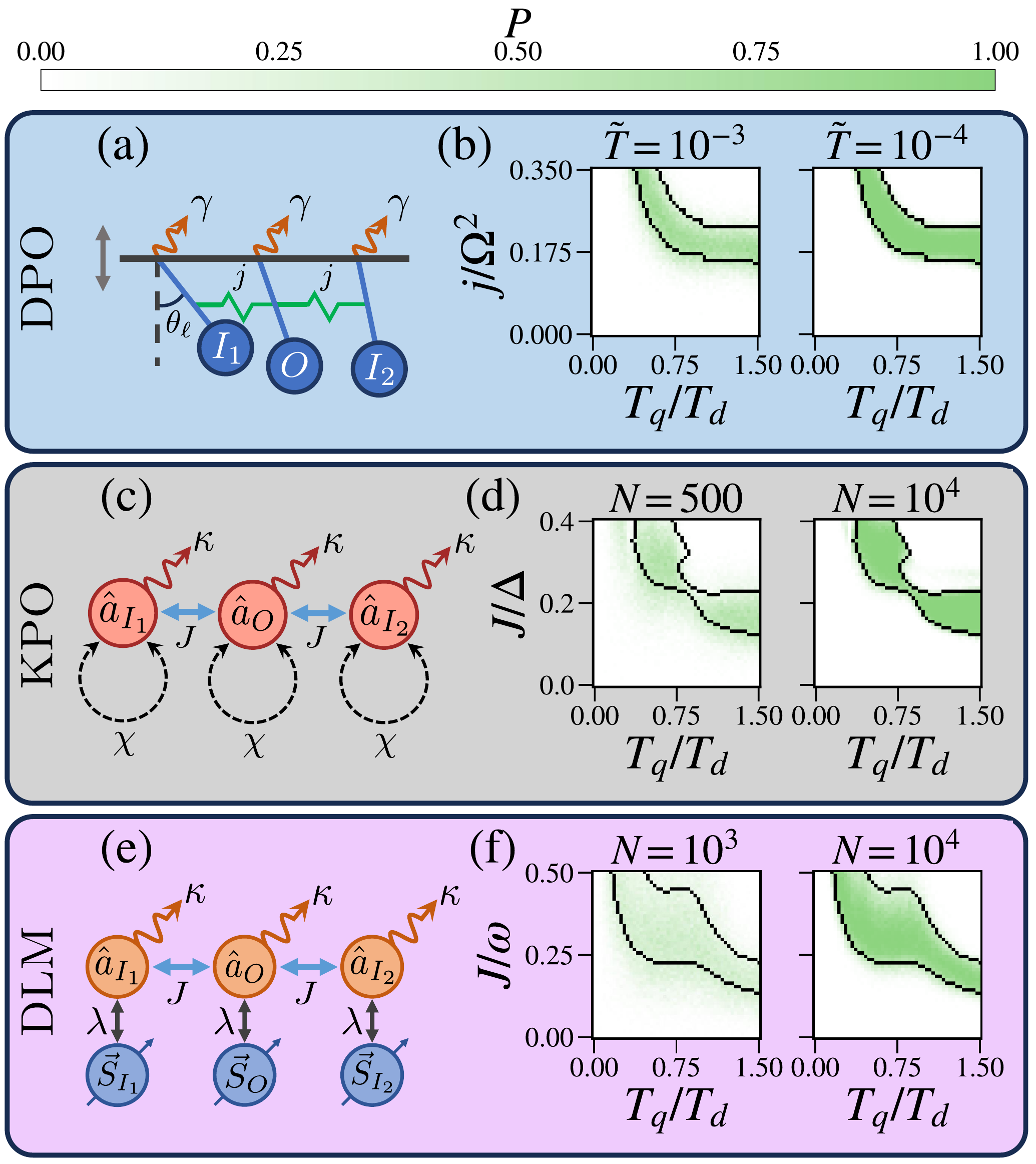}
    \caption{Sketch of the (a) DPO network, (c) KPO network, and (e) DLM. Success probability of the logic gate operation using the symmetry-broken states of the (b) DPO, (d) KPO, and (f) DLM as a function of the $j$, $J$, and $T_q$ for various $N$ and $\tilde{T}$. The black outlines delineates the regions in which $P = 1$ in the noiseless limit ($\tilde{T}=0$ and $N\rightarrow \infty$). The parameters used for the DPO are $\{ \Omega_d, A, \gamma \} = \{2.0\Omega, 0.5, 0.1\Omega \}$. The parameters used for the KPO are $ \{\chi, p_0, A_0, \kappa\} =  \{\Delta, 2.5\Delta, 0.6, 0.4\Delta \}$. The parameters used for the DLM are $ \{\omega_0, \omega_d, \lambda_0, A_1 \} = \{ \omega, 0.8\omega, 0.9\lambda_c, 0.5 \}$.}
    \label{fig:fig3}
\end{figure}

We now examine the effects of thermal noise on the fidelity of our proposed architecture for a DPO network, shown in Fig.~\ref{fig:fig3}(a). We present in Fig.~\ref{fig:fig3}(b) the success probability $P$ of the logic gate architecture for temperatures $\tilde{T} \approx 10^{-3}$ and $\tilde{T} = 10^{-4}$. We calculate $P$ by counting the number of successful operations in 100 different noise realizations. We classify an operation to be successful only if the architecture returns the correct output for all input configurations under the same noise realization. The black outlines corresponds to the boundary of the regions in which $P=1$ in the $T\rightarrow 0$ limit. We note that for finite $T$, the noise can disrupt the operation, causing the logic gate to fail. Noise-induced errors can be attributed to stochastic fluctuations pushing the phase space trajectory into the wrong basin of attraction. This highlights the sensitivity of the protocol when the pulse parameters lead to dynamics close to the basin boundaries. To circumvent this, the pulse operation can be tuned to ensure that the system is driven deeper into the intended basin. As shown in Fig.~\ref{fig:fig3}(b), with decreasing temperature, the high-fidelity region expands to fill the area enclosed by the solid black lines, which represent the $T=0$ prediction.

\textit{Application to other systems --} Having demonstrated that the architecture is valid for coupled DPOs, we verify the applicability of our proposed architecture in open quantum systems described by the Lindblad master equation with single photon loss
\begin{align}\label{eq:lindblad_master_eq}
    \frac{\partial \hat{\rho} }{\partial t} = - \frac{i}{\hbar}  \left[\hat{H}, \hat{\rho} \right] + \kappa (2 \hat{a} \hat{\rho} \hat{a}^\dagger - \frac{1}{2} \{ \hat{a}^\dagger \hat{a}, \hat{\rho} \}).
\end{align}
First, we consider a network of KPOs, shown in Fig.~\ref{fig:fig3}(c), described by the Hamiltonian \cite{yamaji_2023_kpocoupled}
\begin{align}\label{eq:dlm_hamiltonian_hpapprox}
\begin{split}
    \frac{\hat{H}_{\mathrm{KPO}}}{\hbar} &= \sum_{\ell=1}^M \Delta \hat{a}_\ell^{\dagger} \hat{a}_\ell + \frac{\chi}{2} \hat{a}_\ell^{\dagger} \hat{a}_\ell^{\dagger} \hat{a}_\ell\hat{a}_\ell + \frac{p(t)}{2} \Big( (\hat{a}_\ell^{\dagger})^2 + (\hat{a}_\ell)^2 \Big) \\ 
    & -J (\hat{a}^\dagger_\ell \hat{a}_{\ell + 1} + \hat{a}_\ell \hat{a}_{\ell + 1}^\dagger),
\end{split}
\end{align}
where $\chi$ and $p(t)$ are the strengths of the Kerr nonlinearity and two-photon drive, respectively, $J$ is the coupling strength between nearest-neighbor sites, and $\Delta$ is the frequency of the bosonic mode associated with the operators $\hat{a}_\ell$ and $\hat{a}_\ell^\dagger$, which obey $[\hat{a}_\ell, \hat{a}_{\ell'}^\dagger] = \delta_{\ell, \ell'}$. When the two-photon drive is modulated periodically with the form $p(t) = p_0 (1 + A_0 \sin{\omega t})$, the KPO spontaneously chooses between two period-doubled states. 

We next consider the Dicke lattice model (DLM), which describes a system of individual Dicke models arranged on a lattice \cite{zou_2014_dlm}, as depicted in Fig.~\ref{fig:fig3}(e). The corresponding Hamiltonian is given by
\begin{align}\label{eq:DickeLatticeHamiltonian}
\begin{split}
    \frac{\hat{H}_{\mathrm{DLM}}}{\hbar} &= \omega\sum_{\ell=1}^M \hat{a}_\ell^{\dagger} \hat{a}_\ell + \omega_0 \sum_{\ell=1}^M \hat{S}_\ell^z + \frac{2\lambda (t)}{\sqrt{N}} \sum_{\ell=1}^M (\hat{a}_\ell + \hat{a}^{\dagger}_\ell) \hat{S}_\ell^x \\
    &- \sum_{\ell=1}^MJ(\hat{a}^\dagger_\ell \hat{a}_{\ell+1} + \hat{a}_\ell \hat{a}_{\ell+1}^\dagger),
\end{split}
\end{align}
where $\omega$ is the frequency of the bosonic mode, $\omega_0$ is the spin excitation frequency, $\lambda(t)$ is the light-matter coupling, $N$ is the number of spins, $J$ is the coupling strength between nearest-neighbor sites, and $M$ is the number of lattice sites. The operators $\hat{S}_\ell^{x,y,z}$ are the collective spin operators, whereas $\hat{a}_\ell$ and $\hat{a}_\ell^\dagger$ are the bosonic annihilation and creation operators for the $\ell$-th lattice site.

When $\lambda$ is driven periodically with the form $\lambda(t) = \lambda_0 (1 + A_1 \sin{\omega t})$, the Dicke model hosts two symmetry-broken DTC states akin to the period-doubled states of the DPO \cite{nie_2023_dickedtc, zhu_2019_dickedtcs, gong_2018_theoreticaldtc, hans_2021_dickedtc, skulte_2024_experimentaldickedtc}. Here, we consider $\lambda_{0}< \lambda_{c} = \frac{1}{2}\sqrt{\frac{\omega_0}{\omega} (\kappa + \omega^2)}$, where $\lambda_c$ is the critical value for the transition between the normal and superradiant phase for the uncoupled DLM ($J = 0$) \cite{emary_chaos_2003, dimer_proposed_2007, kirton_2019_dickereview}. For this parameter regime, the DLM has an effective mapping to a parametric oscillator as shown in Refs.~\cite{roy_2024_parametricmapping, roy_2025_kzmdtc}.

To simulate the dynamics of the KPO and the DLM networks, we employ a mean-field approximation to probe the dynamics of the two systems in the thermodynamic limit. In this limit, the macroscopic dynamics of the two systems dominate the quantum noise, thus providing a noiseless limit for both systems. To probe the effects of quantum noise, we employ the truncated Wigner approximation (TWA) to incorporate first-order quantum corrections to the systems' dynamics. \cite{polkovnikov_2010_twa, tuquero_2024_twaref}. Additional details regarding TWA and the systems' mean-field equations of motion are included in \cite{suppmat}.

The black outlines in Figs.~\ref{fig:fig3}(d) and (f) enclose the region in which the architecture is valid in the thermodynamic limit for the DLM and KPO. The presence of these regions demonstrate that the protocol applies to both period-doubled systems considered. Using TWA simulations, we obtain the probability of operational logic gates as a function of the coupling strength and the pulse duration in the presence of quantum noise, as shown in Figs.~\ref{fig:fig3}(d) and (f). Similar to the effects of temperature on a network of DPOs, the inherent quantum noise present in the DLM and KPO can decrease the fidelity of the logic gates, as the system operates closer to the quantum regime of few particle number. Increasing the particle number from $N = 10^3$ to $N = 10^4$ brings the system closer to the thermodynamic limit, where the fluctuations become negligible. This brings the probability distributions for both the DLM and KPO closer to the mean-field distribution.

\textit{Conclusion} -- We have provided a simple framework for manipulating the symmetry-breaking states caused by a parametric resonance. Our framework is based on tracking the phase-space trajectory of a measurable quantity, the absolute time phase of an order parameter exhibiting period-doubling oscillation, to identify the coupling strength and duration required for an effective bit-flip operation. We have demonstrated that the same framework can be utilized to create universal logic gates for period-doubling states. We have verified the generality of our proposed logic gate architecture in DLM and KPO networks, and its robustness against either classical or quantum noise. Our work leverages the intricate interplay between dissipation and coupling in systems hosting dynamical states, providing a means for manipulating and encoding information on periodically-driven systems.

\textit{Acknowledgement. - } This work was primarily funded by the Department of Science and Technology (DOST) and monitored by the Philippine Council for Industry, Energy, and Emerging Technology Research and Development (DOST-PCIEERD) with Project No. 1214356.

\nocite{carollo_2021_dickemeanfield, zhangbaranger_2021_kpomeanfield, holstein_1940_hpapprox, seibold_2026_kpotwa}

\bibliography{reference.bib}
\bibliographystyle{apsrev4-2}

\end{document}


\title{Supplemental Material for \\ Universal logic gates for coupled period-doubling systems}

\author{Emmanuel Joseph Lee D.G. U}
\affiliation{National Institute of Physics, University of the Philippines, Diliman, Quezon City 1101, Philippines}
\author{Roy D. Jara Jr.}
\affiliation{National Institute of Physics, University of the Philippines, Diliman, Quezon City 1101, Philippines}
\author{Jayson G. Cosme}
\affiliation{National Institute of Physics, University of the Philippines, Diliman, Quezon City 1101, Philippines}
\date{\today}

\maketitle

\section{METHODS}

\subsection{Mean-field Approximation}

Consider the Heisenberg equation for the expectation value of an arbitrary operator $\hat{O}$,
\begin{equation}
\label{eq:heisenberg_eq}
    \frac{\partial \langle \hat{O} \rangle}{\partial t} = i \left< \left[\frac{\hat{H}}{\hbar}, \hat{O} \right] + \mathcal{D}[\hat{O}] \right>
\end{equation}
where $\hat{H}$ is the Hamiltonian of the system, and $\mathcal{D}[\hat{O}] = \sum_{\ell} \kappa_{\ell} \left( 2\hat{L}_{\ell}^{\dagger} \hat{O} \hat{L}_{\ell} - \left\{ \hat{L}^{\dagger}_{\ell} \hat{L}_{\ell}, \hat{O} \right\} \right)$ is the dissipator associated with the jump operators $\hat{L}_\ell$. In general, Eq.~\ref{eq:master_eq} generates an infinite hierarchy of equations that depend on the moments of $\hat{a}$. To truncate this hierarchy we use the mean-field approximation, valid in the thermodynamic limit $N \rightarrow \infty$ such that correlations can be approximated as $\langle \hat{a} \hat{b} \rangle\approx \langle \hat{a} \rangle \langle \hat{b} \rangle$. This means that the resulting EOM depend only on the first moments of quantum operators, which can be treated as simple complex numbers, $a \equiv \langle \hat{a} \rangle \in \mathbb{C}$.

For a network of KPOs, to initialize the $\ell$-th site as a 0-bit (1-bit), we use the initial state $a_\ell (0) = 0.1$ ($a_\ell (0) = 0.1i$). For the DLM, to initialize the $\ell$-th site as a 0-bit (1-bit), we use the initial states $\{a_\ell (0), b_\ell (0) \} = \{ 0.1, 0.1 \}$ ($\{a_\ell (0), b_\ell (0) \} = \{ 0.1i, 0.1i \}$).

\subsection{Truncated Wigner Approximation}

Consider the master equation for the time evolution of the density matrix,
\begin{equation}
\label{eq:master_eq}
    \frac{\partial \hat{\rho} }{\partial t} = -\frac{i}{\hbar}  \left[\hat{H}, \hat{\rho} \right] + \mathcal{D}[\hat{\rho}],
\end{equation}
where the creation and annihilation operators in the Hamiltonian are bosonic in nature. We employ the TWA to describe the first-order quantum correction to the dynamics of an arbitrary operator $\hat{O}$. To this end, the EOM within TWA is obtained by applying the following operator correspondence,
\begin{equation}
    \hat{a}^\dagger \hat{a} \hat{\rho} \longleftrightarrow \left( \alpha^\ast - \frac{1}{2}\frac{\partial}{\partial \alpha} \right) \left(\alpha + \frac{1}{2}\frac{\partial}{\partial \alpha^\ast} \right)W, \quad \quad
    \hat{\rho} \hat{a}^\dagger \hat{a} \longleftrightarrow W \left( \alpha^\ast + \frac{1}{2}\frac{\overleftarrow{\partial}}{\partial \alpha} \right) \left( \alpha - \frac{1}{2}\frac{\overleftarrow{\partial}}{\partial \alpha^\ast} \right),
\end{equation}
on Eq.~\ref{eq:master_eq}, where $W$ is the Wigner function corresponding to the Wigner-Weyl transform of the density matrix \cite{polkovnikov_2010_twa}. The operator correspondence effectively maps the quantum operators $ \{ \hat{a}, \hat{a}^\dagger, \hat{\rho} \}$ onto the classical phase variables $\{ \alpha, \alpha^*, W(\alpha, \alpha^*) \}$. The resulting equation can be recast into a Fokker-Planck equation
\begin{equation}
\label{eq:fokker-planck}
    \partial_t W = - \sum_{n \in \{ R, I \}} \frac{\partial}{\partial \alpha_n} \left( A_n W \right) + \frac{1}{2} \sum_{m, n \in \{ R, I \}} \frac{\partial}{\partial \alpha_n} \frac{\partial}{\partial \alpha_m} \left[ \left( \mathbf{B}\mathbf{B}^{\mathrm{T}} \right)_{nm} W \right],
\end{equation}
where the diffusion matrix $D = \mathbf{B} \mathbf{B}^\mathrm{T}$ is positive-definite, and the subscripts $\{R, I\}$ denote the real and imaginary parts of $\alpha$. The Langevin equation corresponding to Eq.~\ref{eq:fokker-planck} is given by
\begin{equation}
    \frac{d \mathbf{ \vec{\alpha}}}{dt} = \mathbf{A}(z,t) + \mathbf{B}(z,t) \mathbf{E}(t),
\end{equation}
where $\vec{\alpha} = \left[ \alpha_R, \ \alpha_I \right]^\mathrm{T}$, and the real stochastic term $E_n(t)$ obeys 
\begin{equation}
    \langle  E_n (t)  \rangle = 0, \quad \quad
    \langle  E_n(t) E_m(t')  \rangle = \delta_{nm} \delta(t - t').
\end{equation}
For both the DLM and a network of KPOs, we use the same initialization scheme shown in the discussion for the mean-field approximation, and we suppress the stochastic terms until the pulse operation is to be performed. This is because the noise can randomly push a site into the opposite basin before it has a reached the intended initial steady-state. Such a procedure is equivalent to a post-selection process wherein we seek our desired initial configuration for the network.

\section{Systems of Interest}

\subsection{Kerr Parametric Oscillator}

For a network of coupled KPOs, we use Eq.~\ref{eq:master_eq} to compute for the dynamics of the bosonic field in the $\ell$-th site. We apply the mean-field approximation, which is valid in the large-amplitude semiclassical limit ($|\hat{a}|^2 \to \infty$) for the KPO \cite{zhangbaranger_2021_kpomeanfield}. This yields the EOM
\begin{equation}
\label{eq:eom_kpo}
    i \frac{da_\ell}{dt} = - \Delta a_\ell - \chi a_\ell^* a_\ell^2 + p_0 a_\ell^* - J(a_{\ell - 1} + a_{\ell + 1}) - i\kappa a_\ell
\end{equation}
where we denote $\langle \hat{a} \rangle \equiv a$.

In the mesoscopic limit, the dynamics of the system can be described using TWA. To control the crossover between the quantum and classical regimes, we introduce a tunable parameter that rescales the bosonic field as $a = \sqrt{N} \tilde{a}$. Writing $\tilde{a} \rightarrow a$ for brevity, the EOMs within the TWA are given by \cite{seibold_2026_kpotwa}
\begin{equation}
\label{eq:eom_kpo_twa}
    i \frac{da_\ell}{dt} = - \Delta a_\ell - \chi a_\ell^* a_\ell^2 + p_0 a_\ell^* + i \kappa a_\ell - J(a_{\ell - 1} + a_{\ell + 1}) + \sqrt{\frac{\kappa}{N}} \xi_\ell(t)
\end{equation}
where the fluctuation due to the dissipation is captured by a stochastic term $\xi_\ell(t)$, which satisfies $\left\langle \xi_{\ell'}^* (t) \xi_\ell (t')\right\rangle = \delta_{\ell \ell'}\delta(t - t')$ \cite{seibold_2026_kpotwa}.

\subsection{Dicke Lattice Model}

In order to recast the DLM into a form that readily maps onto the DPO, we apply the Holstein-Primakoff transformation, given by $\hat{S}_{\ell}^{z} = \hat{b}_{\ell}^{\dagger}\hat{b}_{\ell} - \frac{N}{2}$ and $\hat{S}_{\ell}^- = (\hat{S}_{\ell}^+)^\dagger = \sqrt{N} \sqrt{1 - \frac{\hat{b}^\dagger \hat{b}}{N}}\hat{b}$ \cite{holstein_1940_hpapprox, roy_2024_parametricmapping}. We then approximate $S^{-}_{\ell} \approx 1 - \frac{1}{2}\frac{\hat{b}^\dagger \hat{b}}{N}$, which is valid when $\lambda_0 < \lambda_c$, since in the NP $b^{\dagger}b / N \ll 1$. This yields the Hamiltonian
\begin{align}\label{eq:HolsteinPrimakoffHamiltonian}
\begin{split}
  \frac{\hat{H}}{\hbar} &= \omega \sum_{\ell=1}^M \hat{a}_\ell^{\dagger} \hat{a}_\ell + \omega_0 \sum_{\ell=1}^M \hat{b}_\ell^{\dagger} \hat{b}_\ell - \sum_{\ell=1}^MJ_\ell(\hat{a}^\dagger_\ell \hat{a}_{\ell+1} + \hat{a}_\ell \hat{a}_{\ell+1}^\dagger) \\
  &+ \frac{2 \lambda(t)}{\sqrt{N}} \sum_{\ell=1}^M (\hat{a}_\ell + \hat{a}^{\dagger}_\ell) \left(\hat{b}_\ell + \hat{b}^{\dagger}_\ell - \frac{\hat{b}^{\dagger}_\ell(\hat{b}_\ell)^2}{2N} - \frac{(\hat{b}^{\dagger}_\ell)^\ell \hat{b}_\ell}{2N} \right), 
\end{split}
\end{align}
from which the EOM can be derived using Eq.~\ref{eq:heisenberg_eq}. Applying the mean-field approximation, which is exact for the DM in the thermodynamic limit $N \rightarrow \infty$ \cite{carollo_2021_dickemeanfield}, we obtain the following set of EOMs ($\xi_\ell(t) = 0$)
\begin{align}
    i \frac{da_\ell}{dt} &= \omega a_\ell + \lambda(b_\ell + b_\ell^*) \left(1 - \frac{|b_\ell|^2}{2N} \right) - J(a_{\ell + 1} + a_{\ell -1}) -i\kappa a_\ell + \xi_\ell (t),\label{eq:motion_a} \\ 
    i \frac{db_\ell}{dt} &= \omega_0 b_\ell + \lambda(a_\ell + a_\ell^*) \left( 1 - \frac{1}{2N} (2|b_\ell|^2 + b_\ell^2)\right) \label{eq:motion_b}
\end{align}
To include the effects of quantum correlations, we use TWA, which gives rise to the stochastic term $\xi_\ell(t)$ that satisfies $\left\langle \xi_\ell^* (t) \xi_{\ell'} (t')\right\rangle = \kappa \delta_{\ell\ell'}\delta(t - t')$ \cite{tuquero_2024_twaref}.

\section{BASINS OF ATTRACTION}

\begin{figure}[!htpb]
    \centering
    \includegraphics[scale = 0.35]{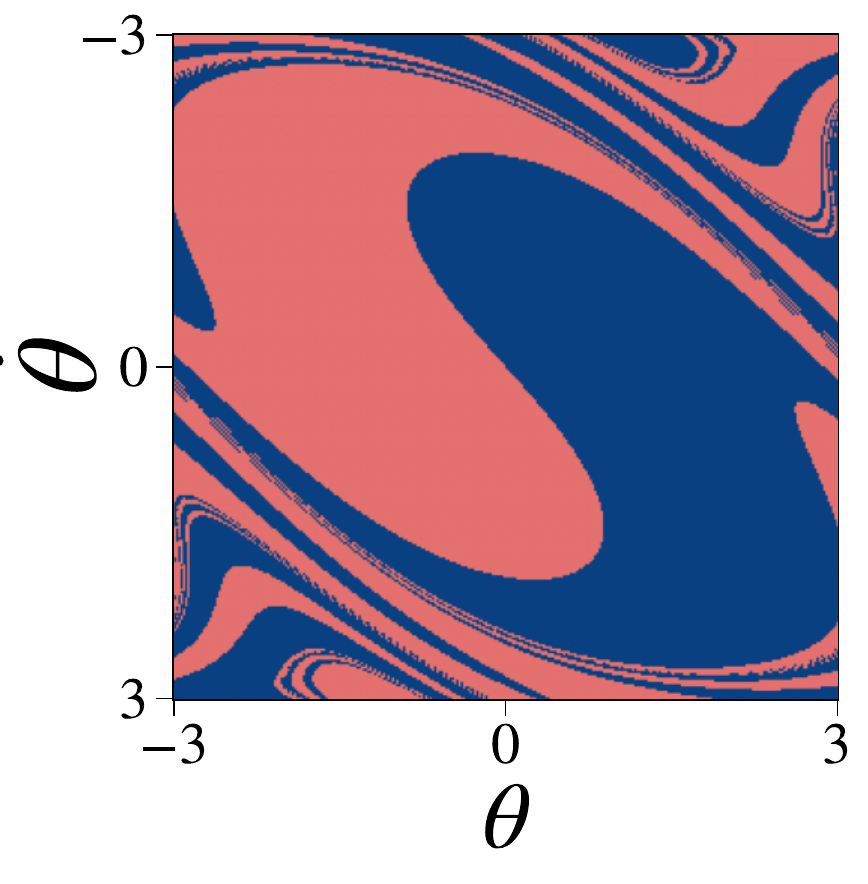}
    \caption{Basins attraction for the two degenerate period-doubled states of the DPO in the lab frame. The parameters used are $\{ \Omega_d, A, \gamma \} = \{ 2.0\Omega, 0.5\Omega, 0.2\Omega \}$.}
    \label{fig:basins}
\end{figure}

In the following, we consider the minimal scenario of a single oscillator in the $\tilde{T} \rightarrow 0$ limit, described by the EOM
\begin{equation}\label{eq:dpo_eom}
    \ddot{\theta} + \Omega^2[1-A\cos (\Omega_d t)]\sin \theta + \gamma \dot{\theta} = 0.
\end{equation}
The state of the system is characterized by the phase-space coordinates $\{ \theta, \dot{\theta}\}$. To construct the basins of attraction for the period-doubled states in the lab frame, we perform a scan over various initial conditions and simulate the dynamics until a time $t_f = 150T_d$, where $T_d = 2\pi / \Omega_d$. We then record the final symmetry-broken state of the system using the absolute time phase $\phi(t_f)$. The results are shown in Fig.~\ref{fig:basins}.

To construct the basins of attraction in the rotating frame, we use the dynamics from $t = 0$ to $t = 2T_d$ to calculate $\phi(t = 0)$, from which we extract the initial state of the system in the rotating-frame phase space spanned by $X = \mathrm{Re}[\phi(t)]$ and $Y = \mathrm{Im}[\phi(t)]$. We again perform a scan over various initial states and use $\phi(t_f)$ to record the final state of the system, yielding the corresponding results in the main text.

\section{Further Details on the Universal Logic Protocol}

\begin{figure}[!htpb]
    \centering
    \includegraphics[scale = 0.45]{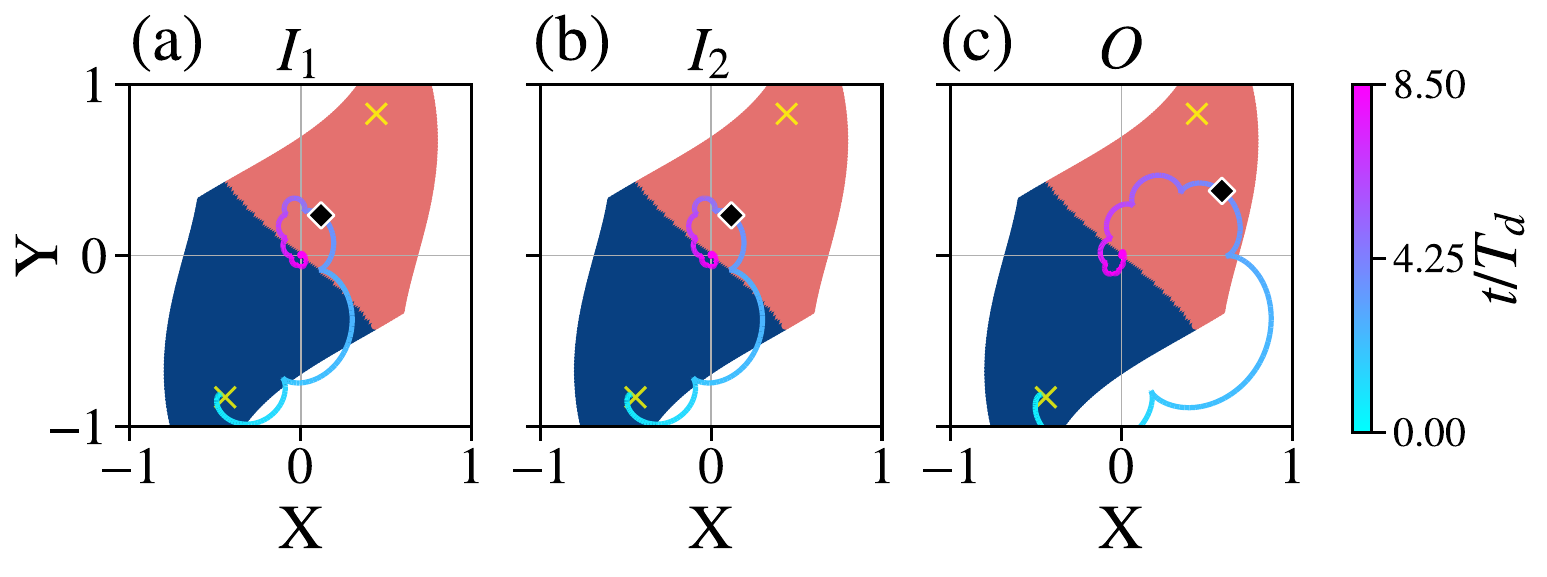}
    \caption{Exemplary dynamics of the protocol calculating the NAND result for $I_1 = I_2 = 1$, showing how the two input sites can be flipped during an operation. The parameters used are $\{  \Omega_d, A, \gamma, j\} = \{ 2.0\Omega, 0.5, 0.2\Omega, 0.3\Omega^2 \}$.}
    \label{fig:input_trajectories}
\end{figure}

\begin{figure}[!htpb]
    \centering
    \includegraphics[scale = 0.2]{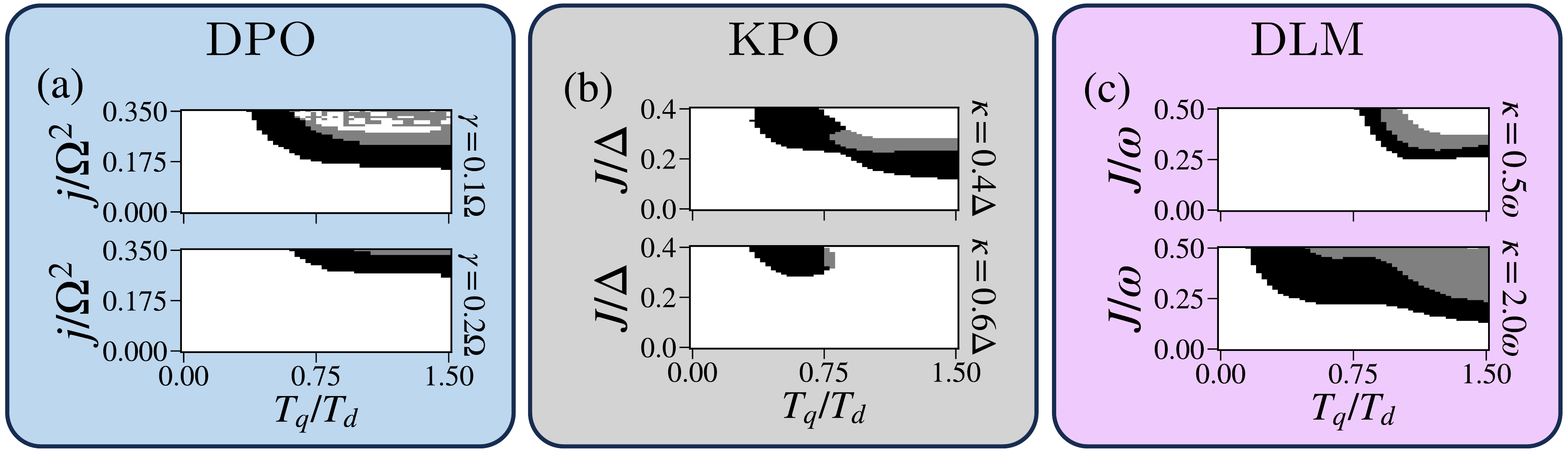}
    \caption{(a) - (c) Success of the logic gates in the noiseless ($\tilde{T} \rightarrow 0$ and $N \rightarrow \infty$) as a function of $j$, $J$, and $T_q$ for the (a) DPO, (b) KPO, and (c) DLM. The black-shaded (gray-shaded) regions indicate when a set of parameters leads to the correct NAND or NOR output for all possible input configurations without flipping the state of any of the input sites (while flipping the state of the input sites for at least one input configuration). The parameters used for the DPO are $\{ \Omega_d, A, \gamma \} = \{2.0\Omega, 0.5, 0.1\Omega \}$. The parameters used for the KPO are $ \{\chi, p_0, A_1, \kappa\} =  \{\Delta, 2.5\Delta, 0.6, 0.4\Delta \}$. The parameters used for the DLM are $ \{\omega_0, \omega_d, \lambda_0, A_1 \} = \{ \omega, 0.8\omega, 0.9\lambda_c, 0.5 \}$.}
    \label{fig:pseudogates}
\end{figure}

We present in Figs.~\ref{fig:input_trajectories}(a)-(c) some representative dynamics of $I_1$ and $I_2$ while coupled to $O$. For the NAND gate, when $I_1 = 1$ and $I_2 = 1$, the coupling can induce switching not only in $O$ but also in the input sites, such as at the point marked with a diamond. This behavior can be undesirable for certain applications of a logic gate, such as for a full-bit adder. Thus, we define a 'pseudo logic gate' as the case in which a given coupling strength and $T_q$ produces the expected behavior for $O$ for all possible input configurations, but flips the input sites for at least one configuration. In Fig.~\ref{fig:pseudogates}(a)-(c), we show the mean-field validity plots for the DPO, KPO, and DLM, respectively. The gray-shaded regions denote a pseudo-logic gate.
We also identify the effect of the dissipation rate on the success of the logic gate. Specifically, increasing $\kappa$ to $\kappa = 2.0\omega$ for the DLM, we observe an increase in the area of the valid region. The opposite is true for the KPO, where we observe a decrease in the area of the valid region when $\kappa$ is increased to $0.6\Delta$, suggesting the existence of a system-dependent optimal dissipation strength leading to the largest area of the valid region in the parameter space.

\section{The Reset Protocol}

\begin{figure}[!htpb]
    \centering
    \includegraphics[scale = 0.19]{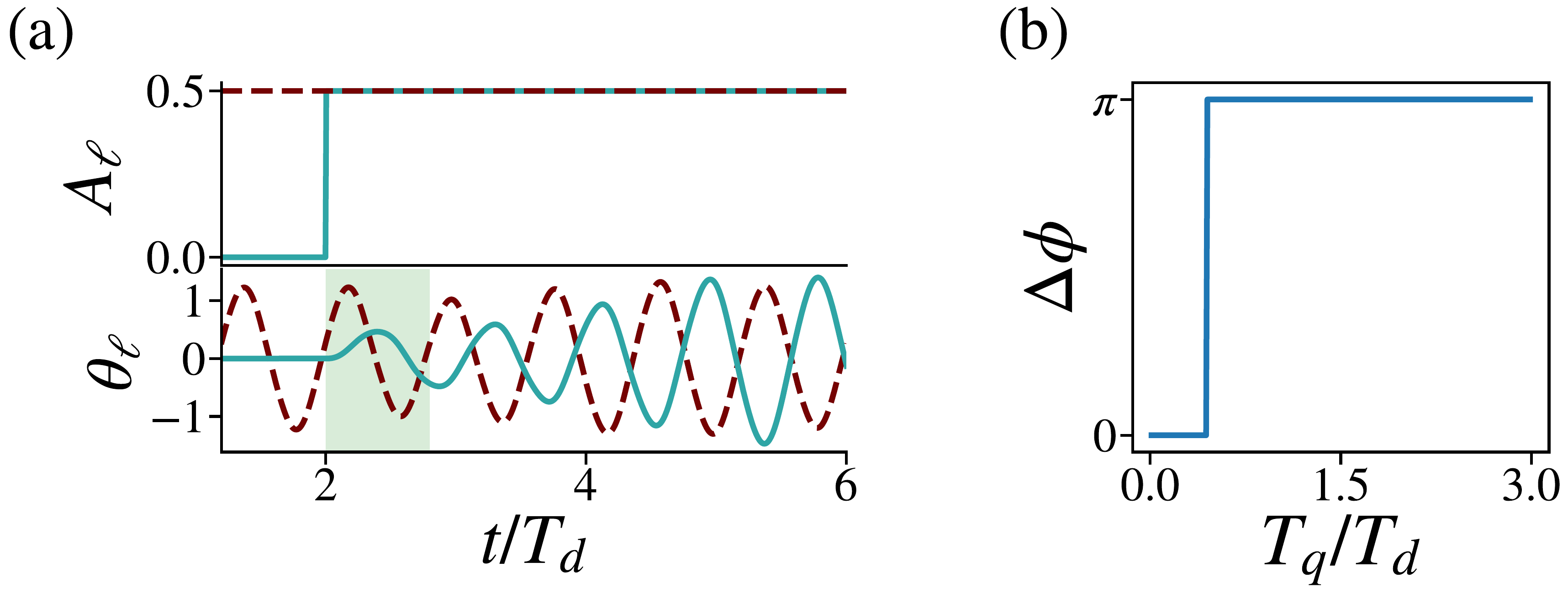}
    \caption{(a) (Top panel) The driving strength $A_\ell$ as a function of time. (Bottom panel) Exemplary dynamics showing the anti-synchronization of the first site (solid) with the period-doubled state in the second site (dashed). The green-shaded region denotes when the coupling is active. (b) $\Delta \phi$ as a function of $T_q$. The parameters used are $\{ \Omega_d, A, \gamma, j\} = \{ 2.0\Omega, 0.5, 0.1\Omega, 0.3\Omega^2 \}$.}
    \label{fig:synch_antisynch}
\end{figure}

We consider a network of two DPOs in which the first site is initially undriven, $A_{\ell = 1} = 0$, and the second site is initialized as a period-doubled state, $A_{\ell = 2} \neq 0$. Then, a pulse is applied to $j$, while at the same time reinstating the periodic driving the first site. The exemplary dynamics in Fig.~\ref{fig:synch_antisynch}(a) show that this sequence causes the first site to anti-synchronize with the period-doubled state in the second site.  More generally, we show in Fig.~\ref{fig:synch_antisynch}(b) that this protocol can result in either synchronization or anti-synchronization with the period-doubled state in the second site depending on the pulse duration $T_q$. To numerically identify the final state of the system, we calculate the quantity $\Delta \phi \equiv | \phi_{\ell = 1} - \phi_{\ell = 2}|$. For synchronization (anti-synchronization), $\Delta \phi = 0$ ($\Delta \phi = \pi$). In the main text, we exploit this behavior to reset the state of the output site after a logic operation, which is required to perform subsequent operations. 

\black
\bibliography{SM_reference}